\documentclass[a4paper, oneside, twocolumn, notitlepage, 10pt]{extarticle_ecoc}
\usepackage{ecoc}

\begin{document}
\selectlanguage{english}    


\title{Experimental Demonstration of Optoelectronic Equalization for Short-reach Transmission with Reservoir Computing}%


\author{
    Stenio M. Ranzini\textsuperscript{(1,2)}, Roman Dischler\textsuperscript{(2)},
    Francesco Da Ros\textsuperscript{(1)}, Henning B\"{u}low\textsuperscript{(2)}, Darko Zibar\textsuperscript{(1)}
}

\maketitle                  


\begin{strip}
 \begin{author_descr}

   \textsuperscript{(1)} DTU Fotonik, Technical University of Denmark, 2800, Kgs.Lyngby, Denmark, \textcolor{blue}{\uline{smara@fotonik.dtu.dk}}

   \textsuperscript{(2)} Nokia Bell Labs, Lorenzstr. 10, 70435, Stuttgart, Germany

 \end{author_descr}
\end{strip}

\setstretch{1.1}


\begin{strip}
  \begin{ecoc_abstract}
    A receiver with shared complexity between optical and digital domains is experimentally demonstrated.  Reservoir computing is used to equalize up to 4 directly-detected optically filtered spectral slices of a 32\;GBd OOK signal over up to 80 km of SMF.
  \end{ecoc_abstract}
\end{strip}


\section{Introduction}

Data-driven technologies are straining the demand for data rates in short-reach communications, i.e., inter- and intra-datacenter. Low-complexity solutions are demanded in those scenarios due to the high amount of transceivers required. In this picture, the direct detection (DD) system is highly attracted given that only a single photodetector (PD) is required in the front-end receiver. Typically the phase information can not be recovered without further effort, which makes chromatic dispersion (CD) a major obstacle in extending the transmission reach.

Solutions to overcome these challenges have already been addressed in the literature in the optical \cite{gruner2000dispersion,litchinitser1997fiber}, digital \cite{mecozzi2016kramers, rubsamen2008isi,jignesh2018transmitter,gaiarin2016high,karanov2019end,yonenaga1997dispersion,hu2019dd} and in both shared domains \cite{secondini2003adaptive,katumba2019neuromorphic,li2019100,argyris2019pam,ranzini2019tunable, da2020reservoir}. The optical solutions focus on compensating all the CD before detection, but a large footprint is required.  Digital solutions rely on reconstructing the phase information from the power received signal \cite{mecozzi2016kramers} and/or applying nonlinear algorithms to attempt to mitigate the distortions\cite{rubsamen2008isi,jignesh2018transmitter,gaiarin2016high,karanov2019end}. Alternatively, there are also solutions to reduce the received signal bandwidth to avoid the spectral power fading that arises from the interaction of CD \cite{yonenaga1997dispersion,hu2019dd}. However, most of them are complex either in the implementation or in the training process to find the optimal set of parameters. Reservoir computing (RC) appears as an alternative to reduce the complexity of the training phase of machine learning algorithms in recurrent topology \cite{jaeger2001echo,maass2002real}. Optoelectronic implementations \cite{vandoorne2014experimental,brunner19PRCbook} of this method have already shown significant gains in transmission reach for optical systems, either in numerical analyses\cite{katumba2019neuromorphic}, or experimentally \cite{li2019100, argyris2019pam}. 

We recently proposed a hybrid optoelectronic approach to leverage on sharing the complexity between optical and electrical domain \cite{ranzini2019tunable, da2020reservoir}. By applying narrow optical filters prior to detection and receiving each spectral slice with a separate PD, we showed that it is possible to recover the sliced information and increased the transmission reach impaired by CD with machine learning algorithms (feedforward neural network and RC). In this work, we experimentally demonstrated an 80-km single-mode fiber (SMF) fiber transmission using the optoelectronic approach with a 32 GBd on-off keying (OOK) signal. The results are compared with a single PD receiver showing a significant improvement in achievable reach.

\begin{figure}[t]
   \centering
        \includegraphics[width=1\linewidth]{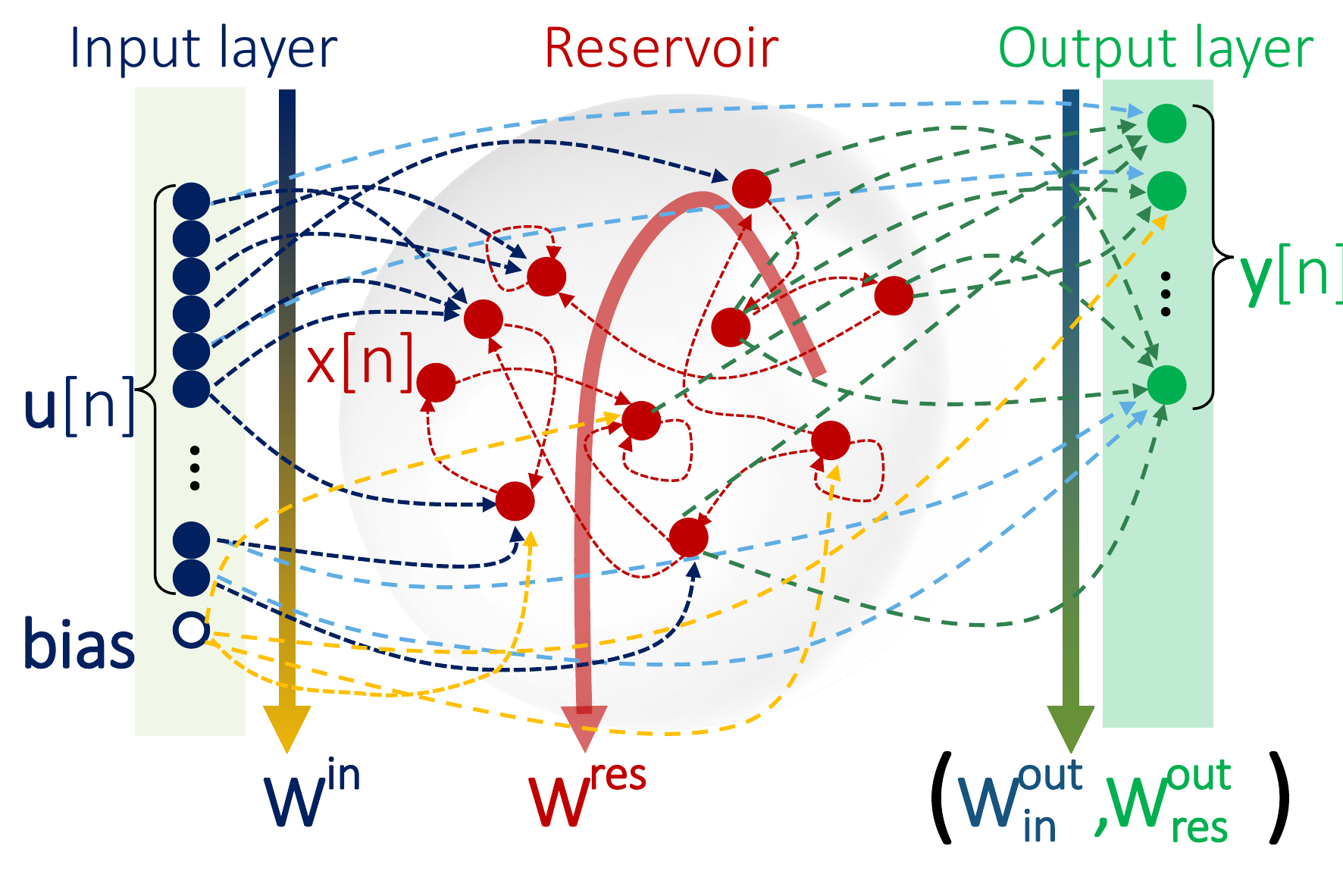}
    \caption{Reservoir computing general structure.}
    \label{fig:RC}
\end{figure}

\begin{figure*}[t]
   \centering
    \includegraphics[width=1\linewidth]{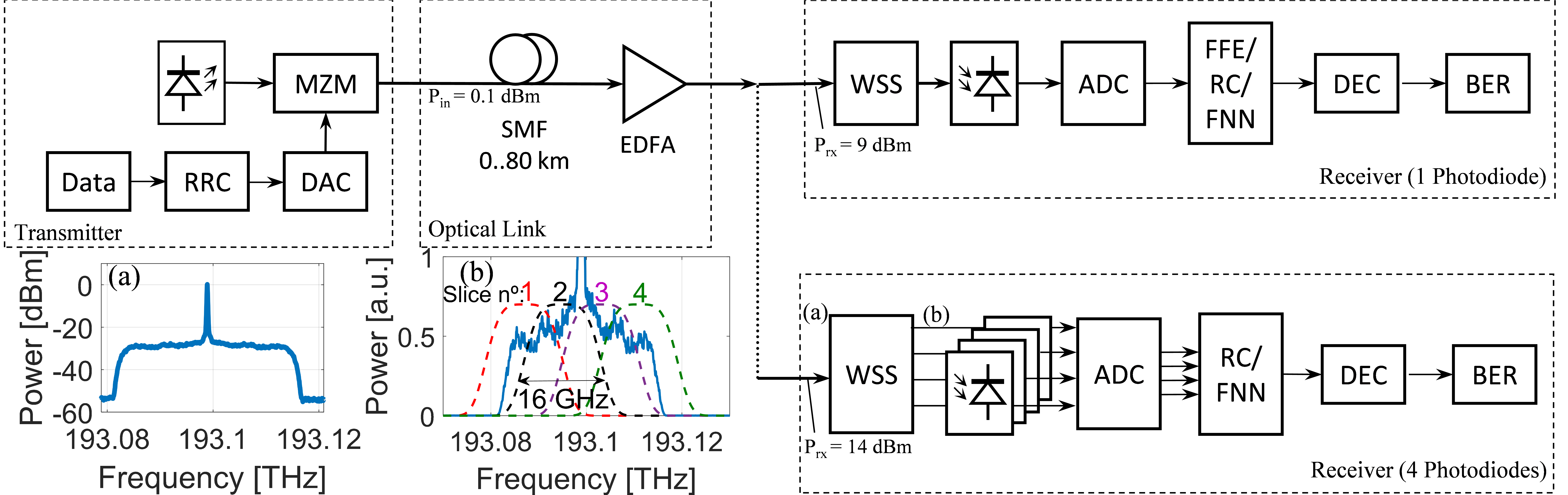}
    \caption{Experimental setup. Inset figures (a) Received spectrum (measured w. 150 MHz resolution bandwidth) after 80 km of SSMF transmission with 6-dB carrier-to-signal power ratio and (b) WSS filters relative position to received spectrum.}
    \label{fig:experimental}
\end{figure*}

\vspace{-2mm}
\section{Reservoir Computing}

Fig.\;\ref{fig:RC} shows the RC general architecture. It is a highly-redundant recurrent neural network that consists of three parts: the input layer, the reservoir and the output layer. The number of inputs in the input layer depends on the implementation. Here, we use one input for each PD (plus bias). The weights connecting input layer to reservoir ($W^{in}$) are drawn from a uniform distribution $\mathcal{U}(-1,1)$. The number of neurons in the reservoir is analyzed in the results section. The probability of interconnections of neurons in the reservoir is drawn from a binary distribution. Additionally, the probability of no interconnection corresponds to 98\% of sparsity. Those non-zeros weights ($W^{res}$) are then defined from a standard normal distribution $\mathcal{N}(0,1)$ and kept fixed. The activation function for the neurons in the reservoir is defined as hyperbolic tangent function. The output layer has 1 neuron with linear activation function and has connections from the input layer ($W^{out}_{in}$) and from the reservoir ($W^{out}_{res}$). These are the only weights that are trained through linear regression.

Equation (\ref{eq:RCx}) describes the reservoir states ($\textbf{x}[n]$). 

\vspace{-8mm}
\begin{multline}
    \mathbf{x}[n] = \alpha\cdot tanh(\mathbf{W}^{in}\cdot \mathbf{u}[n] + \mathbf{W}^{res}\cdot \mathbf{x}[n-1]) + \\
    (1-\alpha)\cdot \mathbf{x}[n-1],
        \label{eq:RCx}
\end{multline}
\vspace{-6mm}

\noindent where the leaking rate $\alpha$ is set to 0.9 and the input neurons are defined by $u[n]$. At the reservoir's output, a linear regression that minimizes  the  square error of the desired signal and $x[n]$ is applied yielding the network's readout $\mathbf{y}[n]$.

\vspace{-2mm}
\section{Experimental setup}

Fig.\;\ref{fig:experimental} shows the experimental setup for optoelectronic equalization. At the transmitter, a random binary sequence at 32 GBd with OOK is generated and shaped by a root-raised-cosine (RRC) filter (roll-off equal to 0.1) at 2 samples per symbol (sps). The resulted signal is resampled to the digital-to-analog converter (DAC) sampling frequency (88 Gsa/s). A Mach-Zehnder modulator (MZM) is then used to modulate the signal with the bias set at the quadrature point. The resulting optical signal is propagated through a length of SMF and amplified by an erbium-doped fiber amplifier (EDFA), irrespective of the equalizer used. Fig.\;\ref{fig:experimental}(a) depicts the received power spectrum after 80 km transmission, showing the optical carrier and the modulated sidebands. At the receiver, the signal is filtered by a wavelength selective switch (WSS). Fig.\;\ref{fig:experimental}(b) shows the sketch of the optical filters for the case with 4 PD and its relative location in the received optical spectrum. The filter's bandwidth choice was based on our numerical analyses in \cite{da2020reservoir}.For the case with 1 PD, the WSS is set to a broadband filter, removing the out-of-band noise. The signal is then detected by a single (or 4) PD with a 3-dB electrical bandwidth of $\approx$\;40 GHz and then  digitally sampled by a real-time scope with 80 Gsa/s and 33 GHz of electrical bandwidth. The received signal is then processed offline, with an anti-aliasing filter, an equalizer and a hard decision (DEC). Specifically, three different equalizers are tested independently, a feedforward equalizer (FFE), RC and a feedforward Neural Network (FNN). Finally, the bit errors are counted, from that BER is evaluated.

The FFE is implemented with 32 taps updated by a least mean square algorithm at 2 sps. The FNN is a 2-layer neural network equalizer with 32 neurons in the hidden layer with hyperbolic tangent as activation function and 1 neuron in the output layer\cite{ranzini2019tunable}. The Levenberg-Marquardt backpropagation algorithm is used to updated the weights at 8 sps with 5\% of the total data used as training. After training, the weights are kept constant. In contrast to the RC, the channel memory needs to be present at the FNN's input layer. In other words, 5 symbols per PD are used (160 neurons plus bias) in the input layer with a window sliding of 1 symbol at the time. The RC is implemented as discussed in the previous section with the same number of sps and training data as the FNN, but operating on only one sample at the time (1 input neuron plus bias).

\vspace{-2mm}
\section{Results}
\label{sec:results}

\begin{figure*}[t]
   \centering
    \includegraphics[width=1\linewidth]{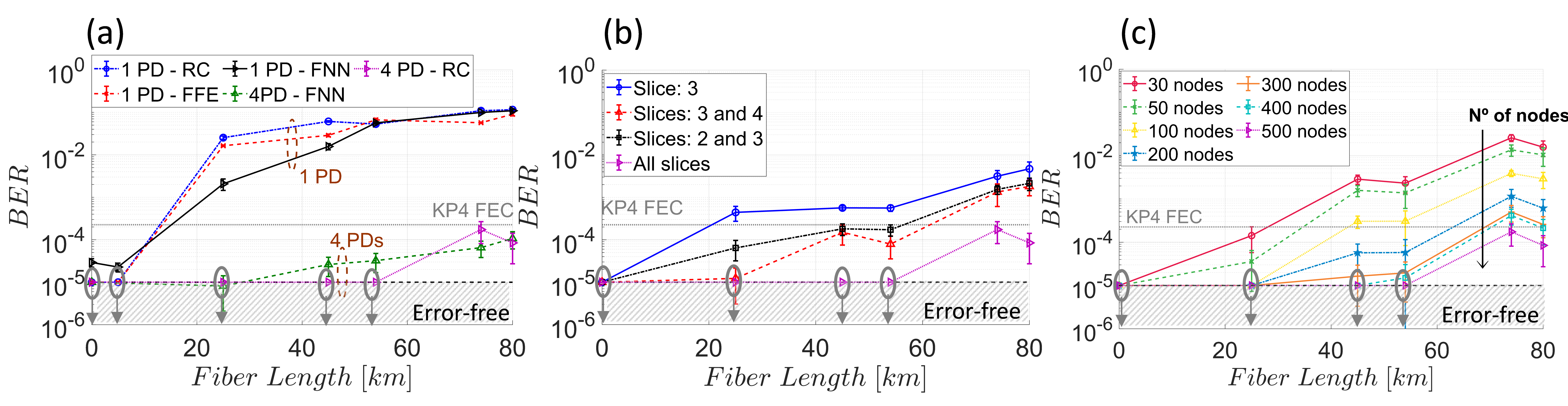}
    \caption{Results. (a) Comparison between 1 PD and 4 PD for different equalizers. (b) Impact of slice's position and number for RC with 500 neurons. (c) Analyses of  increasing the number of neurons in the reservoir.}
    \label{fig:results}
\end{figure*}

Fig.\;\ref{fig:results}(a) shows a comparison of the performance in BER as a function of transmission reach, using 1 and 4 PDs in the receiver for different equalizers. The BER values is calculated over 200k symbols and are averaged over 10 measurements to increase the statistical relevance of the results. The single PD receiver showed a BER below the KP4 forward error correction (FEC) threshold (BER = $2.24\cdot 10^{-4}$ \cite{agrell2018information}) for back-to-back and 5 km fiber transmission. This is expected due to the power fading (spectral notches) in the electrical signal from the interaction between chromatic dispersion and square-law detection, which makes the equalization process challenging. The receiver's digital signal processing fulfills 2 tasks. First is to reconstruct the information from the 4 received signals and second is to mitigate CD. Fig.\;\ref{fig:results}(a) shows that up to 80 km SMF transmission FNN- and RC-equalizers can solve these tasks. However, the differences in the training process and the number of neurons in the input layer should be highlighted. The FNN needs to train all the weights present in the network, while the RC only needs to train the weights in the output layer with a simpler algorithm. The RC's memory comes from the feedback connections and depends on the reservoir properties. Because of that, only 1 sample per PD is necessary for the input layer of RC to mitigate the spreading of the signal over time due to group velocity dispersion of the channel. However, the FNN needs a higher number of inputs in the equalization window to store a similar spreading during the equalization window. 

Note, in the experiment the link loss does not scale linear with link length because different fiber spools with changing connector losses are used. Hence BER, which depends not only on CD but also on receiver OSNR determined by preamp EDFA input power, does not exhibit a monotonically increasing value with length, as can be seen in Fig.\;\ref{fig:results}.


Fig.\;\ref{fig:results}(b) shows the impact of using less slices than the total number available. Using only partial information to recover the transmitted information showed better performance compared to the 1 PD receiver. This indicates the high impact of having notches in the electrical signal's spectrum. The scenario with only the slices 3 and 4 (indicated in Fig.\;\ref{fig:experimental}(b)) can be compared to a single sideband receiver. As well, the slices 2 and 3 can be compared to a duobinary transmission. In both situations with reduced hardware effort, we achieve successful transmission up to 55km with BER below KP4 FEC limit.

Fig.\;\ref{fig:results}(c) shows the impact of number of neurons in the RC. Increasing the number of neurons improved the BER performance. This difference in performance might be related to the memory window available in the reservoir due to the echo state property\cite{da2020reservoir}.  Moreover, increasing the number of neurons also increases the diversity of signals in the RC's output \cite{lukovsevivcius2012practical}. In other words, the dimensionality space of the RC increases, therefore, improving the chances to have a better solution through linear transformation in the output layer to the desired signal.

\vspace{-2mm}
\section{Conclusions}

We showed experimentally an optoelectronic receiver for a transmission with 32 GBd OOK over 80 km of SMF in a DD system. The proposed optoelectronic receiver splits the signal into spectral slices which are then detected separately by PDs. Using all the slices improves reach. Processing of the directly detected spectral slices by different options of NN after ADC showed that RC equalizer and FNN exhibit similar performance. Both equalizers showed a transmission reach of 80 km over the SMF. However, we highlighted the difference in the training process and the number of inputs nodes in the input layer which are more complex for the FNN.  Using a limited number of PDs, 2 instead of available 4, showed degradation in the BER performance. Still, with this reduced hardware effort, a transmission reach of $\approx$ 55\;km below KP4 FEC is demonstrated.

\vspace{-2mm}
\section{Acknowledgements}

This project has received funding from the European Union's Horizon 2020 research and innovation programme under the Marie Sklodowska-Curie grant agreement No 766115, the European Research Council through the ERCCoG FRECOM project (grant No 771878) and VYI OPTIC-AI project (grant No 29344).


\printbibliography

\end{document}